\documentclass[3p,times,procedia]{elsarticle}
\usepackage{nupha_ecrc,amsmath,amsfonts,yfonts,bbm,mathrsfs}

\volume{00}

\firstpage{1}

\journalname{Nuclear Physics A}

\runauth{}

\jid{nupha}

\jnltitlelogo{Nuclear Physics A}

\usepackage{amssymb}

\usepackage[figuresright]{rotating}

\begin{document}

\begin{frontmatter}



\dochead{XXVIIth International Conference on Ultrarelativistic Nucleus-Nucleus Collisions\\ (Quark Matter 2018)}

\title{Entanglement and thermalization}


\author[*]{J\"{u}rgen Berges}
\author[*]{Stefan Floerchinger}
\author[**]{Raju Venugopalan}

\address[*]{Institut f\"{u}r Theoretische Physik, Universit\"{a}t Heidelberg, 69120 Heidelberg, Germany}
\address[**]{Physics Department, Brookhaven National Laboratory, Bldg. 510A, Upton, NY 11973, USA}

\begin{abstract}
In a quantum field theory, apparent thermalization can be a consequence of entanglement as opposed to scatterings. We discuss here how this can help to explain open puzzles such as the success of thermal models in electron-positron collisions. It turns out that an expanding relativistic string described by the Schwinger model (which also underlies the Lund model) has at early times an entanglement entropy that is extensive in rapidity. At these early times, the reduced density operator is of thermal form, with an entanglement temperature $T_\tau=\hbar/(2\pi k_B\tau)$, even in the absence of any scatterings.
\end{abstract}


\end{frontmatter}



\paragraph{Introduction} It is a surprising feature of experimental data on electron-positron collisions, that they show certain thermal-like features. More specific, particle multiplicities can be rather well described by a thermal model using Boltzmann weights with a temperature of about $164.6 \pm 3.0$ MeV \cite{Becattini:2008tx}. This is quite surprising, in particular because conventional thermalization by collisions in the finial state is rather unlikely. Thermal-like features have also been observed in other high-energy collision data and are difficult to understand within \textsc{Pythia}, see ref.\ \cite{Fischer:2016zzs} for a recent discussion. Alternative explanations are needed here. In the following, we will propose a possible explanation, based on intricate features of quantum entanglement in an expanding quantum string. 

\paragraph{QCD strings and entanglement} To start, let us recall that many features of particle production in QCD can be understood within effective string models, for example the Lund model \cite{Andersson:1983ia}. It is believed that the production of quark-antiquark pairs occurs in the expanding string by a version of Schwinger's mechanism and that the mesons can be seen as the fragments of the breaking string. In a quantum formalism, a string is described by a density matrix $\rho$. Local processes in such a situation (within some region $A$ of the string) can then be described by a reduced density matrix, $\rho_A = \text{Tr}_B\{\rho\}$, where the trace goes over the complement region $B$ of $A$. Basically, we will ask here whether there are circumstances under which the reduced density matrix $\rho_A$ resembles a thermal density matrix and how this can affect particle production.  Formally, for a pure quantum state $\rho$, the amount of entanglement between the region $A$ and its complement $B$ can be quantified in terms of the entanglement entropy $S_A = - \text{Tr} \{  \rho_A \ln \rho_A \}$. Interestingly, for a quantum system, a globally pure state with $S=0$ can be locally mixed in the sense that $S_A >0$ as a result of entanglement. The so-called coherent information $I_{B\rangle A}=S_A - S$ can be {\it positive} \cite{NielsenChuang, Wilde}. This is not possible within classical statistics.

\paragraph{'t Hooft model} To describe QCD strings in a quantum model, one might employ the Lagrangian of QCD confined to $d=1+1$ dimensions, i.\ e.\ the 't Hooft model,
\begin{equation}
\mathscr{L} = - \bar \psi_j \gamma^\mu (\partial_\mu - i g { \bf A}_\mu) \psi_j - m_j \bar\psi_j\psi_j - \frac{1}{2}\text{tr}\,  {\bf F}_{\mu\nu} {\bf F}^{\mu\nu}.
\label{eq:tHooftModel}
\end{equation} 
The model employs fermionic fields $\psi_j$ with $j=1,\ldots, N_f$ and also features $\text{SU}(N_c)$ gauge fields ${\bf A}_\mu$ with corresponding field strength tensor ${\bf F}_{\mu\nu}$. Note, however, that there are no propagating gluons in $d=1+1$ dimensions. The gauge coupling $g$ has dimensions of mass and the theory is superficially superrenormalizable. While \eqref{eq:tHooftModel} is a non-trivial interacting theory, that cannot be solved exactly, the spectrum of mesonic excitations is known in the 't Hooft limit $N_c\to \infty$ with $g^2N_c$ fixed \cite{tHooft:1974pnl}. 

\paragraph{Schwinger model} For our purpose, it is interesting to consider a somewhat simplified theory, namely the abelian version of \eqref{eq:tHooftModel}, i.\ e.\ the Schwinger model of QED in $d=1+1$ dimensions. The U$(1)$ charge $q$ has again dimensions of mass and is related to the string tension $\sigma$ via $q=\sqrt{2\sigma}$. Interestingly, for a single fermion $N_f=1$ the theory can be bosonized exactly \cite{Coleman:1975pw}, leading to the action
\begin{equation}
\begin{split}
S = \int d^2 x \sqrt{g} {\bigg \{} & - \frac{1}{2}g^{\mu\nu}\partial_\mu \phi \partial_\nu \phi - V(\phi) {\bigg \} }, \quad\quad\text{with} \quad\quad  V(\phi) =  \frac{q^2}{2\pi} \phi^2  + \frac{m  \,q \, e^\gamma }{2\pi^{3/2}} \cos\left(2\sqrt{\pi}\phi + \theta\right).
\end{split}
\label{eq:bosonizedSchwingerModel}
\end{equation}
For convenience we have directly formulated the model in general coordinates with a metric $g_{\mu\nu}$. The Schwinger bosons $\phi\sim\bar \psi \psi$ correspond to dipoles of fermions and antifermions.  The Schwinger model also features a vacuum angle $\theta$, similar to QCD in $d=3+1$ dimensions. Note that the massless Schwinger model $m=0$ leads to a free bosonic theory and the vacuum angle plays no role there.

\paragraph{Expanding string solution} Let us now consider a highly energetic quark-antiquark pair on trajectories $z=\pm t$ and follow the string that forms between them. It is convenient to formulate the evolution in terms of Bjorken time $\tau=\sqrt{t^2-z^2}$ and to label positions by rapidity $\eta=\text{arctanh}(z/t)$. The metric is then $ds^2=-d\tau^2+\tau^2 d\eta^2$ and there is a useful symmetry with respect to boosts $\eta\to \eta+\Delta\eta$. A quantum state can be specified in a hypersurface of constant $\tau$ by a (field theoretic) density matrix $\rho_\tau$ and a complete description is obtained by following its evolution with $\tau$ (i.\ e.\ from one hypersurface to the next). 

We will now concentrate on the massless Schwinger model, which has a quadratic action and linear field equations. The fermion mass can be neglected in the strong interaction limit of the theory $q\gg m$, but for the expanding string both $m$ and $q$ become actually irrelevant in the limit of early Bjorken times $1/\tau \gg m,q$ as will be discussed in more detail below. 

\paragraph{Gaussian states and entanglement entropy} The string that stretches between the quark-antiquark pair is described by a coherent state with expectation value $\bar \phi(\tau)$. Theories with quadratic action have oftentimes Gaussian density matrices which are fully specified by the expectation values for the field $\bar\phi(\tau)$, its conjugate momentum field $\bar \pi(\tau)$ and connected two-point functions, e.\ g.\ $\langle \phi(\tau,\eta) \phi(\tau, \eta^\prime) \rangle_c$. Moreover, if the full density matrix $\rho$ is Gaussian, this holds also for the reduced density matrix $\rho_A$. The corresponding entanglement entropy can then be written in a rather compact way as 
\begin{equation}
S_A = \frac{1}{2}\text{Tr}_A \{ D \ln D^2 \}, 
\end{equation}
where the trace goes only over the subregion $A$ and the matrix 
\begin{equation}
D(\eta,\eta^\prime) = \begin{pmatrix}
-i \langle \phi(\eta) \pi(\eta^\prime) \rangle_c   &&  i \langle \phi(\eta) \phi(\eta^\prime) \rangle_c  \\
-i \langle \pi(\eta) \pi(\eta^\prime) \rangle_c &&  i \langle \pi(\eta) \phi(\eta^\prime) \rangle_c
\end{pmatrix},
\end{equation}
is linear in connected correlation functions of fields and conjugate momentum fields within $A$ \cite{Berges:2017hne}. Notably, only connected correlation functions appear, so that coherent states that differ only by the field expectation values have equal entanglement properties.

\paragraph{Entanglement in bosonized massless Schwinger model} One my now ask in particular, how a rapidity interval $\Delta\eta$ is entangled with the complement region in an expanding string at some fixed Bjorken time $\tau$. There are different ways to find the answer to this question. One way is to employ a unitary evolution from an interval in rapidity $\eta$ at constant Bjorken time $\tau$ to an interval in laboratory coordinate $z$ at fixed coordinate time $t$ and to use then known results for the entanglement entropy \cite{Berges:2017zws}. Another is to directly work in the expanding coordinate system and to use a quantization in terms of mode functions \cite{Berges:2017hne}. Here we do not repeat the derivations but only discuss the results. First, the entanglement entropy of the rapidity interval $S_A(\Delta\eta, \tau)$ is divergent. This is no surprise and expected for any quantum field theory as a consequence of infinetely many modes (in the UV) that are shared between the regions via the boundaries. More interesting is the entanglement entropy density $\frac{\partial}{\partial\Delta\eta}S_A(\Delta\eta,\tau)$, which is finite. 

The general behavior of this object is a divergence for $\Delta\eta\to 0$ and an exponential decay with $\Delta\eta$ for large interval length $\Delta\eta\gg 1$. This decay is faster for larger dimensionless combinations of the U$(1)$ charge and Bjorken time $q\tau$. An interesting behavior emerges in the opposite, early time limit $q\tau\to 0$. One finds there that the entanglement entropy density develops a plateau, $\frac{\partial}{\partial\Delta\eta}S_A(\Delta\eta, \tau)\to 1/6$. This is remarkable because it implies that the entanglement entropy becomes extensive (it scales with the spatial ``volume'' $\Delta\eta$), which is reminiscent of the entropy in a thermal state.

\paragraph{Conformal limit} In fact, the early time limit can be understood in terms of an emergent conformal symmetry. In the limit where the expansion or ``Hubble'' rate $1/\tau$ is much larger than the mass scales set by $q$ and the fermion mass $m$, the theory becomes effectively free and has a conformal symmetry. One can then derive quite general that $\partial S_A / \partial \Delta\eta=c/6$ where $c$ is the conformal charge. For a massless real scalar field as well as for a massless Dirac field one has $c=1$ (which shows consistency with or without bosonization). For QCD in $d=1+1$ dimensions one would have $c=N_c \times N_f$ and if the two transverse coordinates of the string are taken into account as additional fields (described by the Nambu-Goto action), this would add two more bosonic degrees of freedom. 

\paragraph{Entanglement and temperature} In the early time limit $q\tau\to 0$ one can actually go further and obtain directly the reduced density matrix $\rho_A$ in any rapidity interval $\Delta\eta$. It can be written as $\rho_A = \frac{1}{Z_A}e^{-K}$ where $Z_A=\text{Tr} e^{-K}$ and $K$ is the so-called modular or entanglement Hamiltonian. It is always possible to define such an operator $K$ but here it is in fact a {\it local} expression,
\begin{equation}
K = \int_\Sigma d\Sigma_\mu \, \xi_\nu(x) T^{\mu\nu}(x).
\label{eq:entanglementHamiltonian}
\end{equation}
The integral goes here over the space-time hyper surface of the interval $A$, $T^{\mu\nu}(x)$ is the energy-momentum tensor excitations around the coherent field expectation value, and $\xi^\mu(x)$ is a vector field. Note that \eqref{eq:entanglementHamiltonian} is in fact precisely the form of a thermal density matrix if $\xi^\mu(x)= u^\mu(x)/T(x)$ can be interpreted as ratio of fluid velocity $u^\mu(x)$ and temperature $T(x)$. The situation becomes particularly simple for $\Delta\eta\to \infty$ where $u^\mu$ points simply in $\tau$ direction and the temperature depends only on Bjorken time, 
\begin{equation}
T=\hbar / (2\pi \tau).
\label{eq:temperature}
\end{equation}

\paragraph{Physics picture and conclusions} Technically, the thermal state arises here because two limiting processes do not commute. For any finite value of $q\tau$ one recovers in the limit $\Delta\eta\to \infty$ a pure state with vanishing entanglement entropy density. In contrast, if the early time limit $q\tau \to 0$ is taken first, and $\Delta\eta\to \infty$ afterwards, one obtains a mixed state with extensive entanglement entropy density $\partial S_A / \partial\Delta\eta = c/6$ which can be characterized by the time-dependent temperature \eqref{eq:temperature}. In this early time limit, excitations around the coherent field expectation value appear as entangled pairs of quasi-particles with opposite rapidity wave numbers. There is an influx (and outflux) of such quasi-particles via the boundaries in rapidity which act effectively as a thermal bath. 

It becomes apparent from this discussion that entanglement between different rapidity regions in an expanding QCD string could indeed play an important role. This seems to be the case at early Bjorken times where the expansion rate $1/\tau$ dominates over the string tension scale $q$ and the fermion mass scale $m$. In a next step one must ask what precisely happens when these scales start to become important and whether a thermal spectrum of produced hadrons (with the right temperature) indeed survives in the final state. Finally, it would be highly interesting to investigate the role of quantum entanglement for other areas of heavy ion physics in more detail.

\paragraph{Acknowledgements} This work is part of and supported by the DFG Collaborative Research Centre ``SFB 1225 (ISOQUANT)''. R.~V.'s research is supported by the U.\ S.\ Department of Energy, Office of Science, Office of Nuclear Physics, under contracts No.\ DE-SC0012704.





\bibliographystyle{elsarticle-num}
\bibliography{<your-bib-database>}



\end{document}